\documentclass[pra,superscriptaddress,showpacs,keywords,reprint]{revtex4-1}
\usepackage{graphicx,amsmath,amssymb,amstext,amsfonts,color}
\usepackage{graphicx}
\usepackage{dcolumn}
\usepackage{bm}
\usepackage{amssymb} 
\usepackage{color} 
\usepackage{subfigure}
\usepackage[colorlinks,linkcolor=red,citecolor=blue,urlcolor=blue]{hyperref}
\usepackage{amsmath,amsfonts,amssymb,amstext}

\newcommand{\be}{\begin{equation}}

\newcommand{\ee}{\end{equation}}
\newcommand{\ben}{\begin{eqnarray}}
\newcommand{\een}{\end{eqnarray}}
\newcommand{\bes}{\begin{subequations}}
\newcommand{\ees}{\end{subequations}}
\newcommand{\bF}{\begin{figure}}
\newcommand{\eF}{\end{figure}}

\begin{document}

\title{Comment on ``Entanglement and chaos in the kicked top"}

\author{Vaibhav Madhok}
\affiliation{Department of Mathematics and Department of Zoology,
University of British Columbia, Vancouver, BC V6T 1Z4 Canada}
\email{vmadhok@gmail.com}

\date{\today}

\begin{abstract}
We comment on the investigation of the connection between chaos and dynamically generated entanglement in Phys. Rev. E, \textbf{83}, 016207, (2011). While, in the referred paper, the authors give an explicit example of a state initially localized in the regular region and still has entanglement properties similar to the states localized in the chaotic region, a few clarifications related to previous works are in order.
Firstly, it is crucial to point out that such a behavior can occur for states initially localized near the border between the chaotic region and regular island, which seems to be the case in the example provided in the referred paper. We comment on the time evolution of such states and the extent to which these can be regarded as having ``regular dynamics".
Secondly, the degree to which entanglement is correlated with the chaos in the system is better understood when we analyze the same initial state and increase the chaoticity in the system gradually.
We also discuss in what capacity entanglement can be regarded as a signature of chaos in such studies.

\end{abstract}
    
\maketitle


 The role of dynamical chaos in entanglement generation has been extensively studied in the last decade\cite{fnp, ms99, l01, Ghose, Wang, tmd08, Lakshminarayan/Bandyopadhyay2002, dk04, bl05}. 
Main interest of these works is the connection of chaotic maps with the generation of random states in the Hilbert states that have typically high entanglement \cite{tmd08, Lakshminarayan/Bandyopadhyay2002}. The primary focus here is not whether a particular regular initial condition can generate more entanglement but to study the entangling power of chaotic maps and connections to random matrix theory. 
Therefore, for fully chaotic classical maps, the corresponding quantum counterparts generate random states in the Hilbert space. And the bipartite entanglement for random states is near maximal \cite{tmd08, Lakshminarayan/Bandyopadhyay2002}.
As an example, ror large $d$-dimensional spaces,  the entanglement of a ``typical state" picked at random from a $d_1 \otimes d_2$ tensor product Hilbert space is given by the Haar measure average of the entanglement over the whole space gives \cite{Page, Scott/Caves, Hayden}
\begin{equation}
\bar{E}_{d_1 \otimes d_2}=\sum^{d_1d_2}_{k=d_1+1} \frac{1}{k}-\frac{d_1-1}{2d_2}, d_2\geq d_1.
\end{equation}
For large dimensions, $\bar{E}_{d_1 \otimes d_2} \approx \log d_1 - d_1/(2 d_2)$, which is close to the maximum possible value of entanglement. Fully chaotic dynamics generates typical pure states in an unconstrained bipartite Hilbert space that are highly entangled \cite{Hayden}.
It is in this sense the generation of ``high" entanglement by  chaotic dynamics is \textit{universal}.

  Another observation of these works \cite{tmd08, Wang} is to study the dynamically generated entanglement and its correlation with the classical phase space. Performing numerical calculations of the long-time averaged entanglement generated by the Floquet map, a clear evidence of its relationship to the classical phase space is seen.
The key focus of these studies is the correlation of phase space with entanglement. 
\begin{figure*}
[t]\resizebox{18.1cm}{!}
{\includegraphics{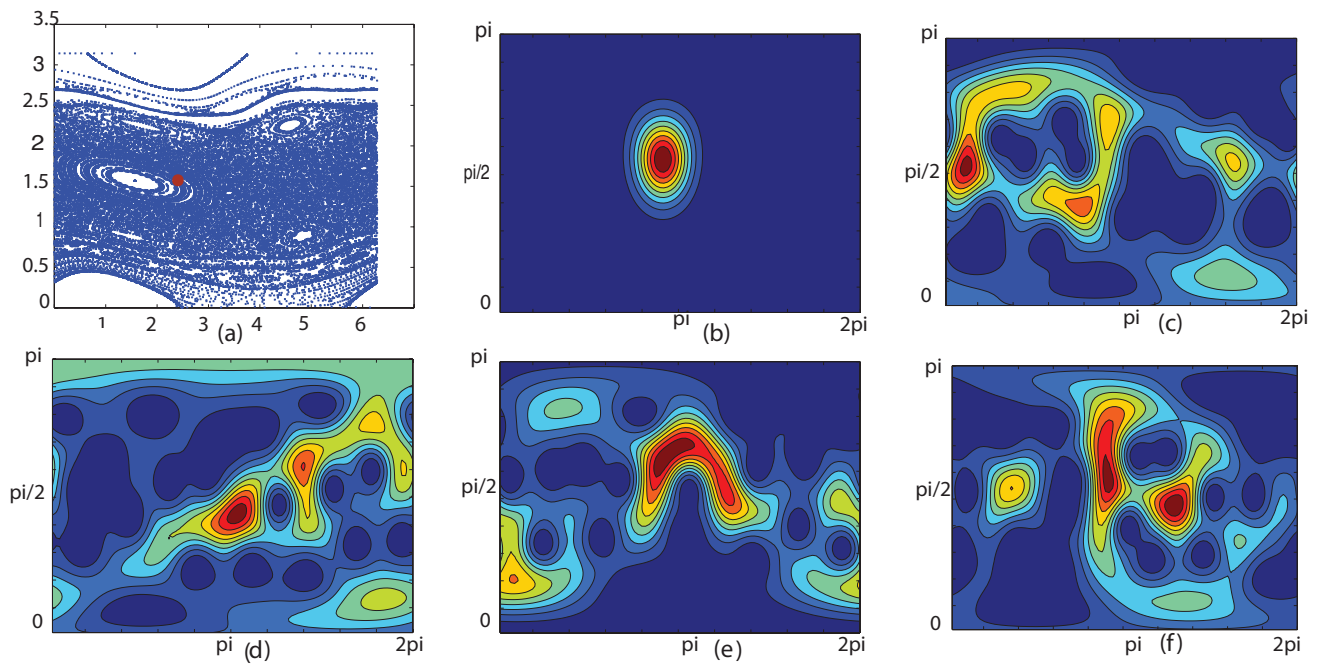}}
\caption{ Flooding of a coherent quantum state, initially localized in the regular region with $\theta=1.7416, \phi=2.8728$ (shown by the red dot in the classical phase space in (a)), in the chaotic sea. (b) At t=1, the Husimi representation of the initial state is well localized.(c) At t = 5, the state is primarily in the regular island with a little support in the chaotic sea. (d)-(f) are the Husimi representations at t=25, t=34 and t=86 respectively, showing the initial state has ``flooded" the chaotic sea. The spin size, $j$, is equal to $10$ and the parameters of the map are $\alpha = 2\pi*0.95$; and $\kappa =5$.}
\label{F8}
\end{figure*}  
Long time average entanglement is higher for states whose initial coordinates are located inside the chaotic sea as compared to states localized deep inside regular islands.
 As we get near the border between a regular island and the chaotic sea ( the ``edge of chaos" in \cite{weinstein}), the time average of the dynamically generated entanglement for an initial condition inside a regular island increases and becomes comparable to a that of a state with chaotic initial conditions. Therefore, entanglement effectively picks out the features of the classical phase space - regular islands from the chaotic sea, while the boundary between the regular island and the chaotic sea remains blurred. 
 While classical chaos can cause infinitely fine structures in the phase space,  the Plank's constant, $\hbar$, limits the scale for such structures in the quantum domain. Our resolution of the phase space is determined by the Plank's constant. 
Therefore, quantum entanglement is a signature of chaos if one considers it as a tool to differentiate between regular islands from the chaotic sea in a coarse grained way. It is in this sense that entanglement is a signature of chaos. This is also the case for quantum maps with large Hilbert space dimension though the difference between average entanglement in the regular region and the chaotic region will be less pronounced for large Hilbert space dimensions.
   
In this regard, for the state corresponding to dot ``2" in Fig 9 in \cite{matzkin}, where 
we are at the border between a regular island and the chaotic sea (but still inside the regular island), the entanglement generated is high and comparable to the state in the chaotic sea ( dot ``1"). 
 Such states are characterized by a high value of entanglement. Such states have similar properties to the states localized in the chaotic sea on other indicators of quantum chaos like the properties of expansion coefficients in the basis of Hamiltonian eigenstates\cite{Zyczkowski}.  It is to be noted that the initial location of the quantum state in the phase space inherently has some information about the system dynamics as can be seen in the expansion of such a state in the basis of Floquet eigenstates and its subsequent time evolution in this basis. The entropy of overlap of these states with the Floquet eigenstates is a signature of chaos \cite{Zyczkowski}.

 It is important analyze the system in a way it gives us an unambiguous correlation of entanglement with chaos.  Matzkin et al. pick a particular state on the border between chaos and regular island and compare its entanglement properties to another state in the chaotic sea. In their approach the dynamics (i.e. the quantum map) remains the same while they compare two different \textit{initial conditions}. They do find that regular initial conditions can generate more entanglement.
However, to what degree ``regular initial" conditions is equivalent to a ``regular dynamics" for the quantum map when we have a map that classically generates a mixed phase space in unclear.
Quantum dynamics, like classical dynamics, has two important components - initial conditions and the time evolution. It is unclear
how a regular initial condition for a quantum map whose classical counterpart has a mixed phase space (or even an almost chaotic phase space in the extreme case!) can be considered and classified as  ``regular dynamics".
 For systems with a mixed phase space the
separation into regular and chaotic ``dynamics" is complicated and chaotic eigenstates ``flood" in the regular islands \cite{bkm05}. From a dynamical point of view, the wave packet from the regular islands can flood into the chaotic sea and vice versa (Fig 1). Therefore,  such a dynamics can hardly be regarded as ``regular" dynamics and the high dynamical generation of entanglement is not surprising.  Moreover, the comparison of this with a ``classically regular dynamics" is of limited value.

Such a dynamics wont have the necessary eigenvalue statistics and level repulsion which is a hall mark of regular quantum maps and it is inaccurate to classify such a dynamics as ``regular". Fully regular quantum maps corresponding to a completely regular classical classical dynamics have Poisson level statistics which gets perturbed when we consider mixed phase spaces. The example given in \cite{matzkin} has a ``regular" initial conditions, however the map as a whole generates intermediate level of chaos through a mixed phase space.
It is for this reason when we talk about entanglement as a signature of chaos, we either consider entangling power of fully chaotic maps or the association of phase space with entanglement in a ``coarse grained" fashion. Therefore, we disagree with this being regarded as a ``refutation" of universality \cite{matzkin2} as, to the best of our knowledge,
this was never the claim in the previous works.

 Moreover, this distinction gets even more blurred when we consider the so called ``edge states", i.e the state on the border between a regular island and the chaotic sea \cite{weinstein} as we have already discussed.


\begin{figure}
\includegraphics[width=8cm, height=5cm]{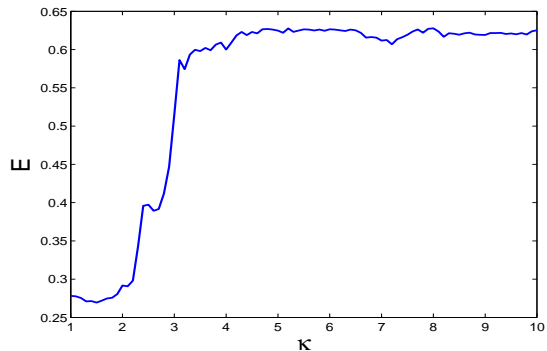}
\caption{Average entanglement, $E$, as a function of the chaoticity parameter, $\kappa$, for the kicked top for $j = 10$. The average is calculated over the first 350 kicks.
The initial state is given by $\theta=2.25, \phi=1.05$.}
\label{F8}
\end{figure}  
 It is therefore important to see how chaos correlates with the dynamically generated entanglement for the \textit{same} initial state as we increase the chaoticity parameter in a systematic manner. The rationale for this is to study unambiguously how, for a fixed family of maps, the degree of chaoticity is related to entanglement generation for a particular coherent states.  We can then determine how does this initial state behave under different dynamics characterized by different degrees of chaos. 
Another way is to average the entanglement over initial states in the regular island and compare it with the average value of entanglement for initial states over chaotic sea.

   We give an example of this approach.
We consider the quantum kicked top \cite{Haake, Wang} and take the initial states as coherent states in the familiar way.   To study the dynamics of entanglement, we use an N-qubit representation of the QKT \cite{Ghose, matzkin},  we trace out two qubits and calculate entanglement between these two qubits and the rest of the system. The entanglement measure chosen is the linear entropy. 
Fig. 2  shows the average entanglement generated for
the same initial state, given by ($\theta=2.25, \phi=1.05$) and $j=10$, as we gradually increase the chaoticity parameter. The average is calculated over the first 350 kicks. The number of kicks chosen should be long enough to capture the periodic modulations of entanglement in the regular region.
We see a strong correlation between the degree of chaos in the system and the average value of entanglement generated.
As we increase the degree of chaos in the system, the initial state transitions from being in the regular island, to be on the border and then finally lies in the chaotic sea.  Therefore, for a fixed family of maps, the generation of entanglement is correlated with the degree of chaos in the system. 
We believe that the initial states mentioned in \cite{matzkin} follow similar entanglement behavior as a function of chaoticity. 
In order to study the entanglement generated in different parts of the phase space, 
we calculate the average entanglement of the chaotic and 
regular regions, we take a grid of coherent states across the 
phase space. 
Each point on the grid is classified as as “regular” 
or “chaotic” by the Lyapunov exponent of the classical 
dynamics. Weighting these values according to the measure on 
phase space gives us an average entanglement of $E = 0.615$ in the 
chaotic sea and $E = 0.344$ in the regular islands. Therefore, 
using the average value of entanglement, one can distinguish regular 
islands from the chaotic sea. In \cite{Wang} it was shown how the average entanglement contour plots
correlate with the classical phase space (Fig. 1 and Fig. 6). This approach helps us to see a quantitative difference between the entanglement generated in different regions of phase space without considering a particular state. 

To summarize, we have briefly discussed in what way entanglement is regarded as a ``universal" signature of chaos
and this comment is to clarify and complement the findings of \cite{matzkin}.


\end{document}